\documentclass{PoS}
\usepackage{graphicx}
\usepackage{rotating}
\title{X-ray selected Narrow-Line Seyfert 1 Galaxies}

\ShortTitle{X-ray selected NLS1s}

\author{{A. Caccianiga}\\
        INAF-Osservatorio Astronomico di Brera, Milan, Italy\\
        E-mail: \email{alessandro.caccianiga@brera.inaf.it}}

\author{P. Severgnini, R. Della Ceca, A. Corral, R. Fanali, 
E. Marchese\\
         INAF-Osservatorio Astronomico di Brera, Milan, Italy\\
         }

\abstract{
We present and discuss the properties of a complete and well defined 
sample of X-ray 
selected type~1 AGNs including 26 Narrow-Line Seyfert 1 and 129 
Broad-Line 
Seyfert 1 galaxies derived from the 
XMM-Newton Bright Serendipitous Survey (XBS). We study the differences 
between the two classes of AGNs for what concerns the X-ray properties
(the photon-index) and the main physical parameters, like the Black-Hole mass 
and  the Eddington ratio. We then consider the two classes together and look 
for 
statistical correlations between observed and physical parameters. 
We find a significant dependence of the photon-index with the Eddington
ratio and a weaker (possibly secondary) correlation between the photon 
index and the Black-Hole Mass.}

\FullConference{Narrow-Line Seyfert 1 Galaxies and their place in the 
Universe - NLS1,\\
		April 04-06, 2011\\
		Milan Italy}

\begin{document}

\section{Introduction}
Narrow-Line Seyfert 1 (NLS1s) Galaxies have been identified as a peculiar AGN 
sub-class on the basis of their relatively narrow Balmer lines 
(\cite{Osterbrock et al. 1985}) and, in some cases, strong optical Fe II 
lines (e.g. \cite{Veron-Cetty et al. 2001}). 
Subsequently, it has been shown that
NLS1 are statistically different from Broad-Line Seyfert 1 (BLS1) galaxies 
also for 
what concerns their X-ray properties, having - on average - steeper photon 
indices and a stronger soft-excess (e.g. \cite{Leighly et al. 1999}, 
\cite{George et al. 2000}, \cite{Grupe et al. 2004}, 
\cite{Bianchi et al. 2009}), \cite{Grupe et al. 2010}). 
Considering the global class of type~1 AGN (NLS1+BLS1) 
a strict anti-correlation between the X-ray photon-index and the H$\beta$ 
width has been observed, with NLS1 occupying one extreme of the distribution
(\cite{Boller et al. 1996}, \cite{Laor et al. 1997}, \cite{Grupe et al. 2004}, 
\cite{Shemmer et al. 2008},
\cite{Risaliti et al. 2009}, \cite{Grupe et al. 2010}). 
The current interpretation of this result is that the physical driver of the 
correlation is
the normalized accretion-rate, i.e. the Eddington ratio (L/L$_{Edd}$), although
it cannot be excluded that other parameters (like the Black-Hole Mass, 
M$_{BH}$) 
can play an important role (\cite{Porquet et al. 2004}, 
\cite{Piconcelli et al. 2005}). Clearly, the study of these 
correlations can shed light on the physical mechanisms  at work in the inner 
part of an AGN, like the disk-corona connection (e.g. \cite{Ishibashi et 
al. 2010}). 

Statistically complete and well-defined samples of type~1 AGNs are 
instrumental to study these fundamental correlations and their
physical implications. The critical requirement of these samples, however, is
to contain enough information, both in the optical and in the X-rays, to 
allow the computation of the main physical parameters (M$_{BH}$, L/L$_{Edd}$),
from the one hand, and to carry out a reliable  X-ray spectral analysis for 
all the sources, on the other hand.
The joint availability of hard X-ray data, from XMM-{\it Newton} and Chandra, 
and of statistical relations that allow the systematic computation of M$_{BH}$ 
on large numbers of AGNs, has produced in the very recent years 
a big leap forward on this 
kind of study, allowing to extend the analysis on significantly
larger samples, including up to a few hundreds of sources.
Some of these studies are  based on the exploitation of the 
large XMM-{\it Newton} database, concentrating either on the targets of the 
XMM-{\it Newton} observations (e.g. \cite{Bianchi et al. 2009}, 
\cite{Zhou et al. 2010},  
\cite{Shemmer et al. 2008}) or on the serendipitous XMM-{\it Newton} 
detections of 
AGN included in the SDSS DR5 spectroscopic database 
(\cite{Risaliti et al. 2009}). 
Usually, these studies are not based on purely flux-limited samples so 
the selection effects cannot be fully controlled. 

We present here the analysis of the type~1 AGNs selected in the 
XMM-{\it Newton} Bright Serendipitous (XBS) survey 
(\cite{Della Ceca et al. 2004}). 
The XBS is a flux-limited survey with a very high identification 
(spectroscopic) level, something that makes its statistical exploitation 
highly reliable. 
Thanks to the relatively high flux limit of the survey all the sources are
detected with a number of net-counts that is reasonably large to perform a
reliable X-ray spectral analysis. 
The analysis of all the $\sim$300 AGN has been already concluded and 
published (\cite{Corral et al. 2011}) and we can now study the 
statistical dependences of the photon index (computed in the XMM-{\it Newton} 
0.5-12 keV energy range) with other physical parameters. 
In Section~2 we present the sample and the classification criteria adopted
to separate NLS1 from BLS1. We also briefly outline the procedure used to 
estimate of the physical parameters (M$_{BH}$ and L/L$_{Edd}$).
In Section~3 we study the statistical correlations observed between 
the computed photon-index and both the Black-Hole Mass and the Eddington ratio.
The conclusions are summarized in Section~4.

\section{NLS1 and BLS1 in the XBS survey}
The XMM-{\it Newton} Bright Serendipitous Survey (XBS survey, 
\cite{Della Ceca et al. 2004})
is a wide-angle ($\sim$28 sq. deg) high Galactic latitude ($|b|>$20 $\deg$) 
survey based on the XMM-{\it Newton} 
archival data. It is composed of two samples both flux-limited 
($\sim$7$\times$10$^{-14}$ erg cm$^{-2}$ s$^{-1}$) in two separate energy 
bands: the  0.5-4.5 keV band (the Bright Serendipitous Sample, BSS) 
and the ``hard'' 4.5-7.5 keV 
band (the Hard Bright Serendipitous Sample, HBSS). 
A total of 237 (211 for the HBSS sample) 
independent fields have been used to select   400 sources, 
389 belonging to the BSS sample and 67 to the HBSS sample (56 sources are
in common). The details on the fields selection strategy, the source
selection criteria and the general properties of the 400 objects 
are discussed in \cite{Della Ceca et al. 2004}. 
To date,  the spectroscopic identification level has reached 
92\% but, for what concerns the type~1 AGN, the identification 
level is  close to 100\%. 
The results of the spectroscopic campaigns are discussed in 
\cite{Caccianiga et al. 2007}, \cite{Caccianiga et al. 2008}. 

\begin{figure}[!t]
\begin{center}
\includegraphics[angle=0,scale=0.5]{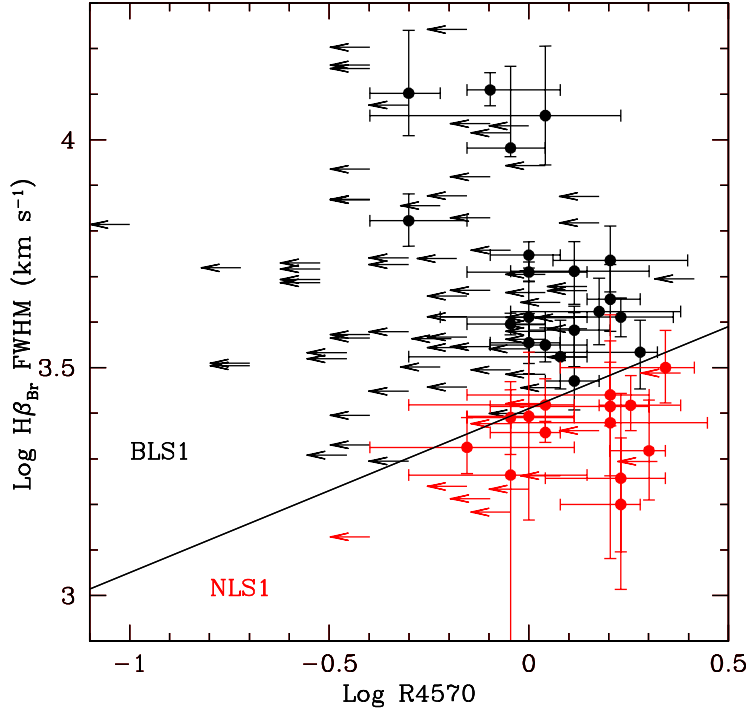}
\caption{Diagnostic plot used to separate NLS1 from BLS1 in
the XBS survey based on the strength of the Fe4570A in
respect to the H$\beta$ total flux (R4570) and the width of the H$\beta$
broad component. Arrows represent upper limits (iron lines not clearly 
detected). The dividing line is taken from \cite{Veron-Cetty et al. 
2001}. Given the
wavelength coverage of our spectra, this classification tool
can be applied only to the AGNs with z$<$0.8.}
\label{diagnostic}
\end{center}
\end{figure}

In total, the XBS sample contains about 300 AGN out of which 
$\sim$170 have a redshift low enough (below $\sim$0.8) to allow the sampling
of the spectral region around the H$\beta$  line. This region is important
to correctly classify an AGN as NLS1. We adopt here the classification 
criterion
suggested by \cite{Veron-Cetty et al. 2001} which is based on the
strenght of the Fe4570\AA\ in respect to the H$\beta$ total (broad+narrow) 
flux (the R4570 parameter) and the width of the H$\beta$ broad component.
According to the authors, this classification criterion is physically
more meaningful to separate NLS1 from BLS1 in respect to a method
based on the H$\beta$ width alone. In any case, it is clear also from the 
analysis
of the XBS sample that type~1 AGN are uniformly distributed in the
R4570/H$\beta$ FWHM parameter space so a separation between NLS1 and BLS1
is always somewhat arbitrary.   
To compute these quantities we have first subtracted from the spectrum
an iron template and fitted the resulting data with a model
composed by 4 Gaussians plus a PL continuum. Two Gaussians are used to 
model the narrow and the broad H$\beta$ line and two are used to 
model the 2 narrow [OIII] lines. 
In Fig.~\ref{diagnostic} we report the R4570 and the H$\beta$ width 
for all the type~1 AGN of the sample together with the dividing line
proposed by  \cite{Veron-Cetty et al. 2001} to separate
NLS1 from BLS1. Using these criteria, we have found 26 NLS1 (26 in the 
BSS and 7 in the HBSS) and 129 BLS1. NLS1 represent about 17\% (1$\sigma$ 
range [14\%-20\%]) of the 
AGN1 in the XBS for which the classification can be applied 
(i.e. with z$<$0.8).
No significant difference in terms of fraction of NLS1 between
BSS and HBSS has been found, considering the statistical errors. 

For all the NLS1 and BLS1 of the survey we have determined both the 
fundamental physical parameters (Black-Hole Masses and Eddington ratios)
and the X-ray properties. A few details are given below.
\begin{figure}[!t]
\begin{center}
\includegraphics[angle=0,scale=0.35]{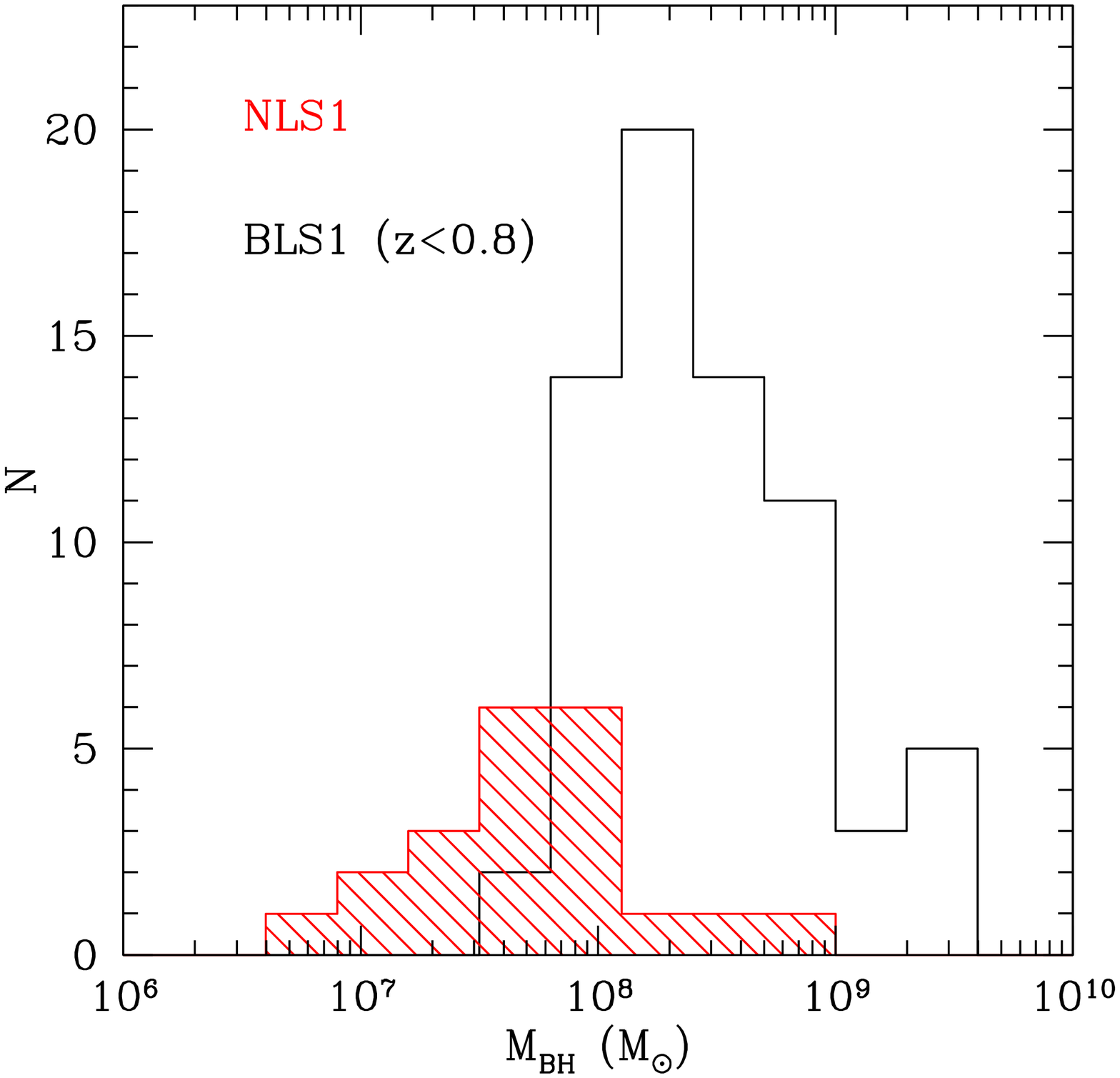}
\includegraphics[angle=0,scale=0.35]{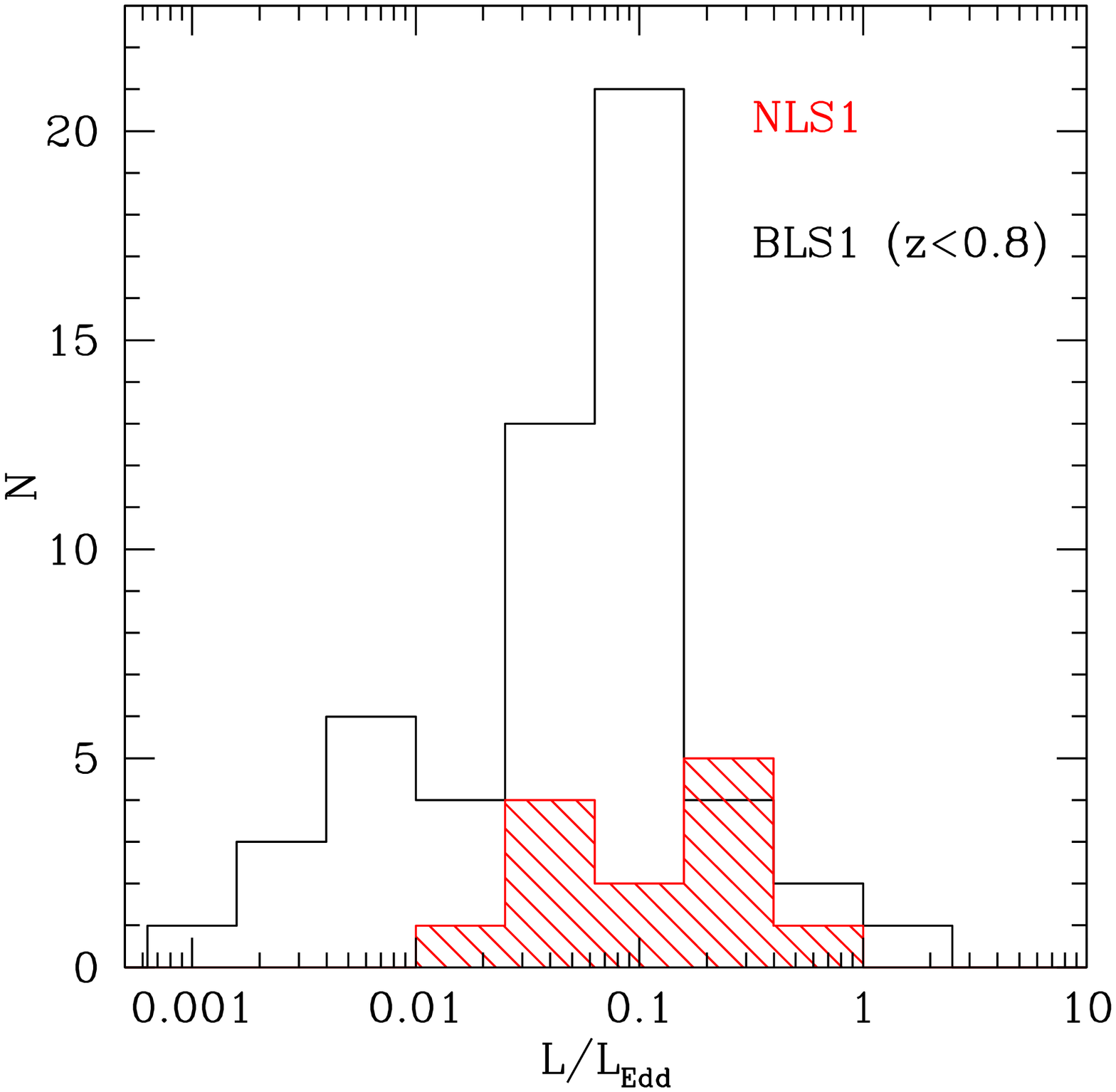}
\includegraphics[angle=0,scale=0.35]{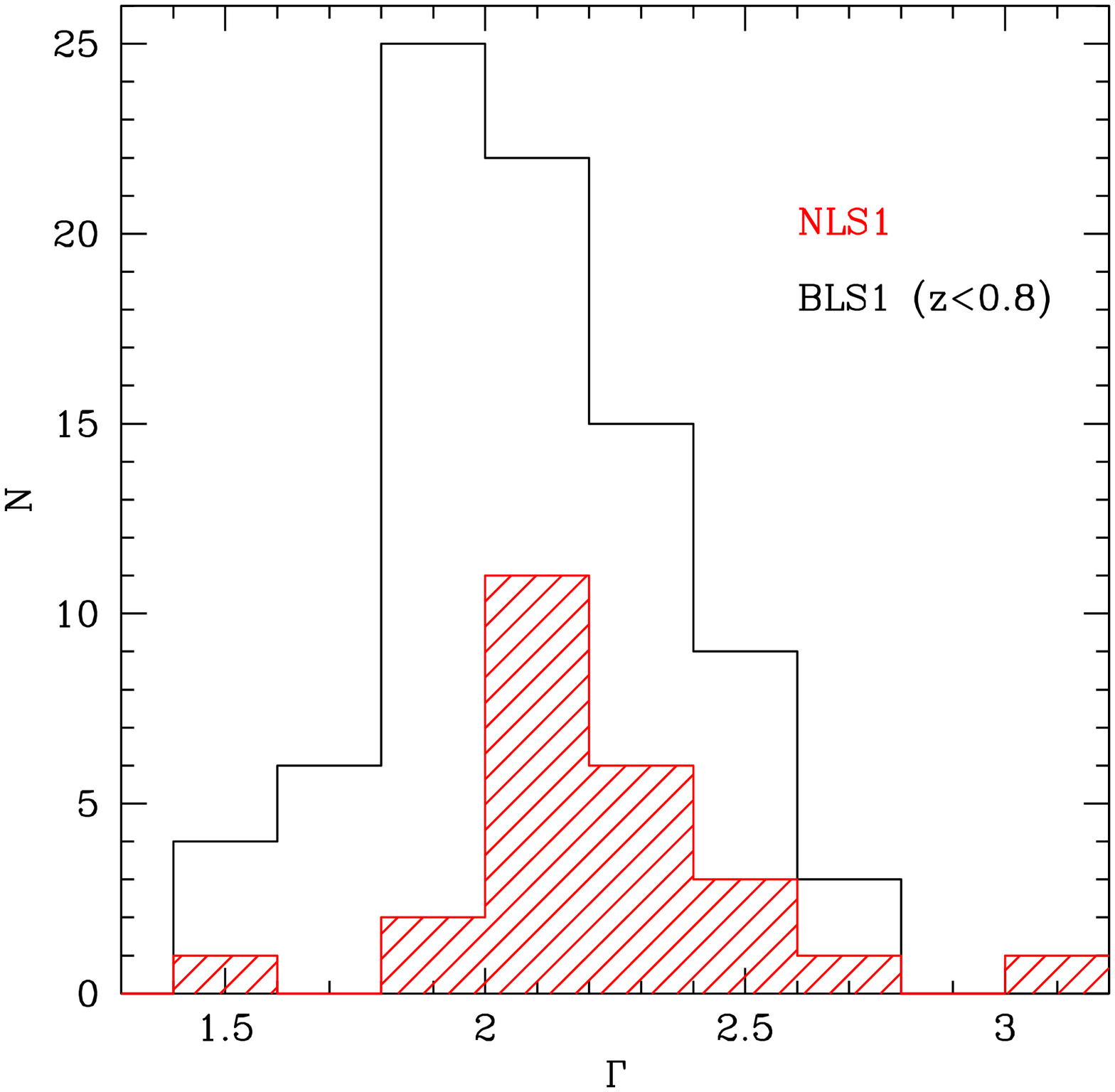}
\caption{Comparison between BLS1 and NLS1 in the XBs survey 
in terms of 
single epoch Black-Hole Mass ({\bf top-left}), Eddington-ratio 
({\bf top-right}) and $\Gamma$ ({\bf bottom}). 
}
\label{distr}
\end{center}
\end{figure}

{\it Black-Hole Mass} - 
We have estimated the Black-Hole 
Masses using the broad H$\beta$ component and applying the
recipes discussed in \cite{Vestergaard et al. 2006}. No correction for the 
radiation pressure has been applied. As expected, the M$_{BH}$ distribution 
of NLS1 and BLS1 (Fig.~\ref{distr}, top-left) are significantly 
(KS D=0.71, probability$<$1\%) different. 

{\it Eddington-ratio} - To compute the Eddington-ratios we have used the 
bolometric luminosities calculated by \cite{Marchese et al. 2011} by fitting
the optical/UV Spectral Energy Distributions built using the available
UV (GALEX) and optical data. Again, the two distributions of Eddington-ratio
of NLS1 and BLS1 are statistically different (K-S test D=0.47, 
probability$\sim$1\%, Fig.~\ref{distr}, top-right).

{\it Photon-index} - 
A systematic X-ray analysis has been carried out on all the AGN of the XBS
survey. The X-ray spectra have been fitted using, as basic model, an 
absorbed power-law plus, in some cases, additional components like
a soft excess, an emission line or a reflected component (see \cite{Corral 
et al. 2011} for details). 
In Fig.~\ref{distr} (bottom) we show the distribution of the photon indices 
computed for the BLS1 and NLS1 (with z$<$0.8) separately. The two
distributions are statistically different (K-S test D=0.32, 
probability = 2.7\%)
although we clearly observe a large spread of values in both samples: 
even if the NLS1 preferentially have steep ($>$2) spectral 
indices, we still observe a small number of flat indices. At the
same time, many BLS1 show steep indices similar to those observed in NLS1. 
The presence of relatively
``flat'' NLS1s has been already pointed out in the literature (e.g. 
\cite{Zhou et al. 2010}).

\section{X-ray spectral index vs. physical parameters: the role of NLS1s}
\begin{figure}[!t]
\begin{center}
\includegraphics[angle=0,scale=0.35]{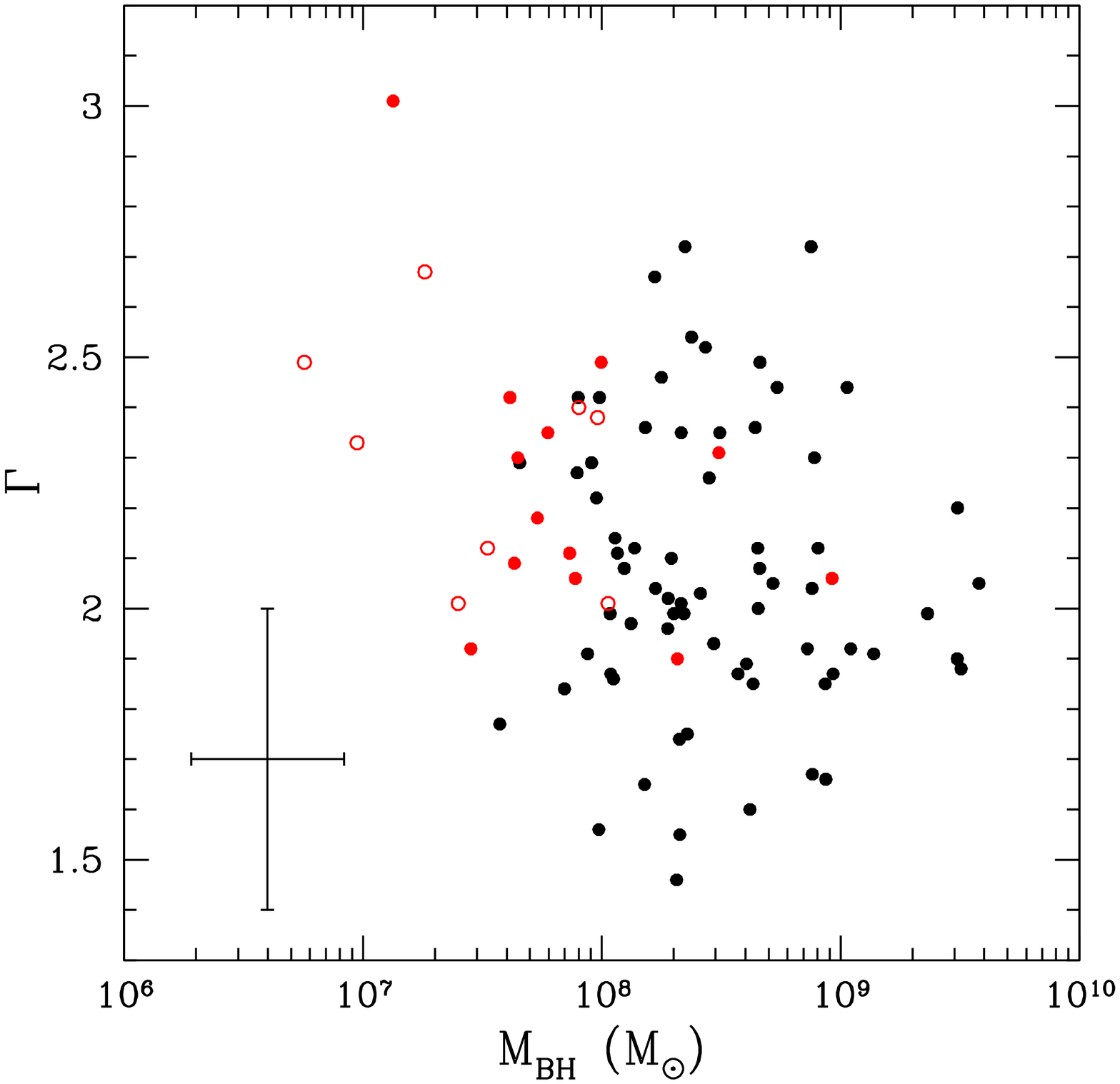}
\includegraphics[angle=0,scale=0.35]{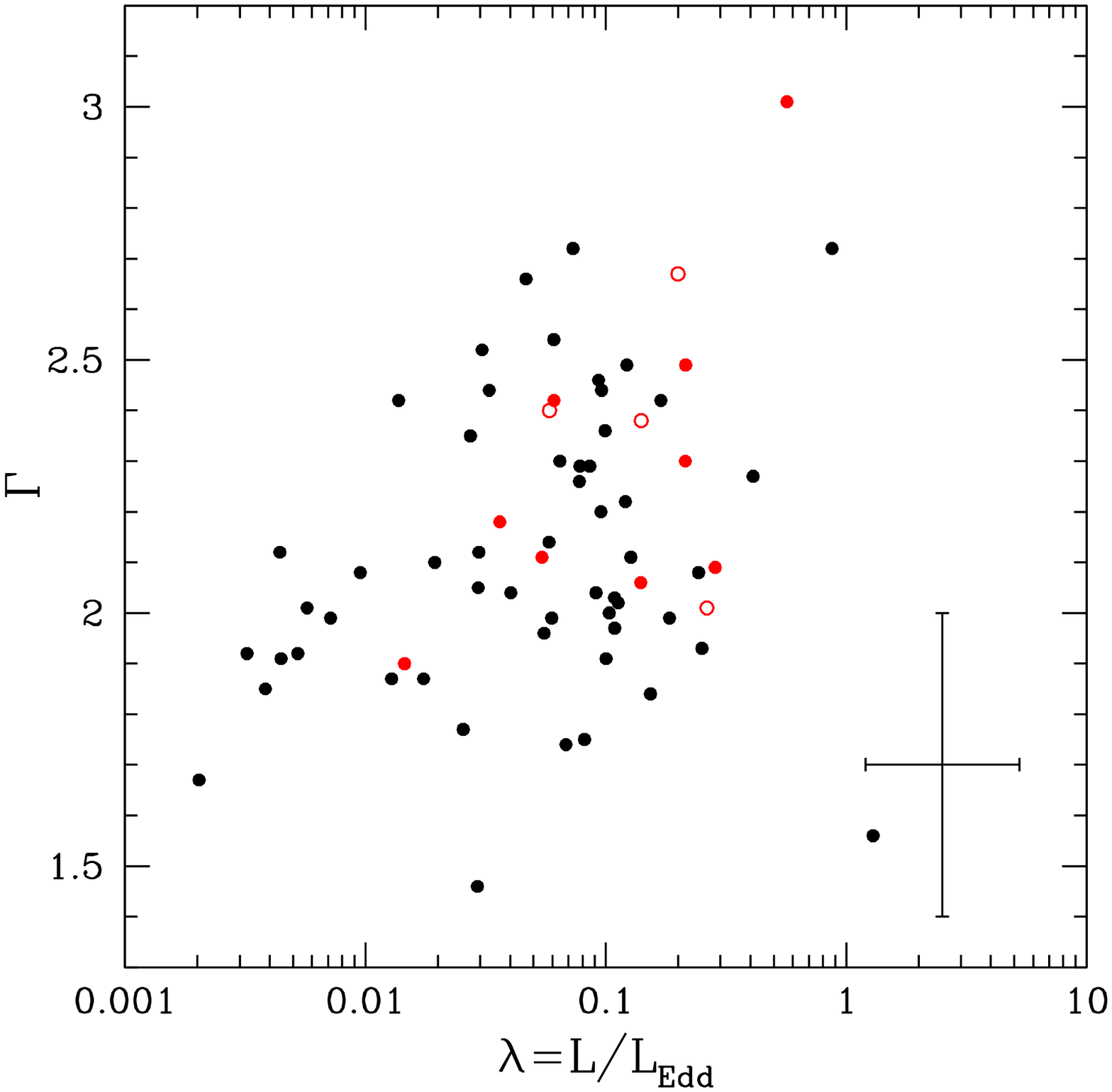}
\caption{Photon index versus M$_{BH}$ ({\bf left}) and 
vs L/L$_{Edd}$ ({\bf right}). Black points are BLS1 while red points are NLS1
(open circles represent uncertain classifications).
Typical error bars are indicated.}
\label{corr}
\end{center}
\end{figure}

In Fig.~\ref{corr} we show the values of $\Gamma$ versus the 
estimated values of M$_{BH}$ and L/L$_{Edd}$ while, in Tab.~\ref{correlation}, 
we report the significance of the correlations. 
For a small number of
objects the computation of L/L$_{Edd}$ was not possible because of the
lack of UV data.
A significant ($>$95\%)/highly significant ($>$99\%) 
anti-correlation/correlation 
is found between $\Gamma$  and  M$_{BH}$ and L/L$_{Edd}$ 
respectively.
We have tested the importance of the NLS1 in the observed correlations, 
by excluding them from the analysis. If we consider just the BLS1 the only
existing correlation is the one between $\Gamma$ and L/L$_{edd}$ although 
the significance is marginal ($>$90\%). The fact that $\Gamma$/M$_{BH}$ 
correlation
disappears when we consider only the BLS1 is probably related to the
fact that the lowest values of M$_{BH}$ ($<$5$\times$10$^{7}$ M$\odot$) 
are covered almost exclusively by the NLS1 and, therefore, their exclusion
limits significantly the dynamic range of the distribution. This is not the
case for what concerns the L/L$_{Edd}$ ratio whose dynamic range is not
significantly affected by the exclusion of NLS1s.

Since M$_{BH}$ and L/L$_{Edd}$ are clearly coupled, it is possible that 
the two observed correlations are not independent. In order to 
establish if this is the case  and in order to assess 
which is the primary correlation, we have run the Spearman
rank test between $\Gamma$ and one of the two parameters, by excluding the
dependence of the third one (see Tab.~\ref{correlation}). 
While the $\Gamma$--L/L$_{edd}$ correlation
remains significant even after the exclusion of the dependence on M$_{BH}$,
the $\Gamma$--M$_{BH}$ correlation strongly weakens when the dependence on 
L/L$_{edd}$ is considered. We conclude that the $\Gamma$--L/L$_{edd}$
correlation is probably the primary dependence while the $\Gamma$--M$_{BH}$
could be just a secondary one. These results are consistent with recent
studies based on XMM-Newton or Swift-XRT data (\cite{Shemmer et al. 2008},  
\cite{Risaliti et al. 2009}, \cite{Grupe et al. 2010}). 

A similar correlation between the soft (E$<$2.4 keV) $\Gamma$ and L/L$_{Edd}$
has been found a few years ago by \cite{Laor et al. 1997} and 
\cite{Grupe et al. 2004} using ROSAT data.
Since the soft X-rays are potentially contaminated by the soft-excess
a possible interpretation of the observed correlation was that 
the soft-excess, and not the intrinsic slope of the X-ray spectrum, was 
related to the Eddington ratio. The fact that this correlation is now 
confirmed 
also using hard (0.5-12 keV) X-ray data, where the contamination from the 
soft-excess is 
lower, supports the idea that it is the intrinsic spectral slope 
that correlates
with the Eddington ratio. The physical implications of this result are many.
It is possible, for instance, that the large number of
seeds photons from the accretion disk, connected to a high accretion rate, 
are responsible for a more efficient cooling of the corona electrons thus 
producing a steeper X-ray spectrum, as recently suggested by  
\cite{Ishibashi et al. 2010}. 
 
%
\begin{table}

\label{correlation}      
\centering                          
\begin{tabular}{c c c c c}        
\hline                
                       & $\Gamma$ vs M$_{BH}$ & $\Gamma$ vs L/L$_{Edd}$ &  $\Gamma$ vs M$_{BH}$   & $\Gamma$ vs L/L$_{Edd}$ \\
                       &                      &                         & (excluding L/L$_{Edd}$) &  (excluding M$_{BH}$)   \\
                       &        (1)           &         (2)             &          (3)            &          (4)            \\
\hline                 
BLS1+NLS1              &  significant         &  highly significant     &  no correlation         &  significant            \\
                       &   ($>$95\%)          &    ($>$99\%)            &                         &   ($>$95\%)             \\
BLS1                   &   no correlation     &    marginal             &  no correlation         &   marginal              \\
                       &                      &    ($>$90\%)            &                         &    ($>$90\%)            \\ 
\hline                            
\end{tabular}
\caption{Results of the correlation analysis (Spearman rank test). Columns 3 and 4
report the results of the partial correlation analysis used to exclude from the
correlation between 2 quantities the hidden dependence on a third variable}%
\end{table}

\section{Summary and conclusions}
We have presented  the properties of a sample of 26 NLS1 and
129 BLS1 extracted from the XBS survey that has a complete optical and X-ray 
spectral characterization. 
We have compared the X-ray photon-indices, derived from the 
XMM-{\it Newton} data, of NLS1 and BLS1 and searched for possible correlations
between $\Gamma$ and the fundamental physical parameters, like the Black-Hole
Mass and the Eddington ratio. The conclusions can be summarized as follows:

\begin{itemize}

\item The NLS1/AGN1 fraction in our X-ray selected sample is ~17\% 
(1$\sigma$=[14\%-20\%]), 
taking into
account the limit of applicability of the classification criteria (z<0.8). 
We do not observe any significant dependence of this fraction with the X-ray 
selection band (0.5-4.5 keV vs. 4.5-7.5 keV);

\item The NLS1 are characterized by a photon-index which is (on average)
steeper when compared to BLS1, by a higher Eddington ratio and by a smaller 
Black-Hole Mass;
 
\item We find that the X-ray spectral index significantly 
correlates with the Eddington ratio. This result is in line with recent 
studies based on XMM-Newton/Swift-XRT data (\cite{Shemmer et al. 2008}, 
\cite{Risaliti et al. 2009}, \cite{Grupe et al. 2010})
and with similar results based on the analysis of the soft ($<$2keV) photon 
index (\cite{Laor et al. 1997}, \cite{Grupe et al. 2004}). 
The fact that NLS1s have, 
on average, steeper X-ray spectral indices than
BLS1 is probably the consequence of their (on-average) high values of 
L/L$_{Edd}$;

\item We find another possible correlation of $\Gamma$ with M$_{BH}$ 
although the
partial correlation analysis shows that this is likely a secondary correlation
induced by the $\Gamma$-L/L$_{Edd}$ dependence.

\end{itemize}

\section*{Acknowledgments}
The authors acknowledge financial support from ASI (grant n. I/088/06/0, 
COFIS contract and grant n. I/009/10/0).

\end{document}